\documentclass[conference,a4paper,9pt]{IEEEtran}
\usepackage{cite}
\usepackage{amsmath,amssymb,amsfonts}
\usepackage{graphicx}
\usepackage{textcomp}
\usepackage{xcolor}
\usepackage{bm}
\usepackage{mathtools}
\usepackage{url}
\usepackage{booktabs}
\usepackage{tikz}
\usetikzlibrary{arrows.meta,positioning,fit,calc,backgrounds,shapes.geometric}

\usepackage[margin=1.7cm]{geometry}

\graphicspath{{figures/}}

\newcommand{\yh}{\hat{y}}
\newcommand{\zh}{\hat{z}}
\newcommand{\cat}{\mathbin\Vert}

\begin{document}
% \ninept

\title{Opportunistic Conditional Entropy Coding with\\ Frozen Analysis and Synthesis Transforms
}

\author{\IEEEauthorblockN{Vincent Corlay, Maxime Rousselot, and Andriy Enttsel}
\IEEEauthorblockA{Mitsubishi Electric R\&D Centre Europe}
}

% \name{Vincent Corlay, Maxime Rousselot, and Andriy Enttsel}
% \address{Mitsubishi Electric R\&D Centre Europe, Rennes, France}

\maketitle

% \begin{abstract}
% Learned image codecs are being standardised, which fixes the analysis and
% synthesis transforms and makes any conditional-coding extension that redesigns
% them impractical. In several delivery settings a lower-quality or
% lower-resolution version of the image has already reached the receiver and could
% serve as side information, but that side information is incidental: it may not be
% there at all, and when it is, its quality is whatever an earlier transmission
% happened to use. Existing conditional codecs instead assume one designed
% condition that is always present. We therefore train a single \emph{opportunistic}
% entropy model. A switch lets it consume the previously decoded latent when one
% exists and fall back to the ordinary hyperprior when none does, while the
% analysis and synthesis transforms stay frozen. Conditioning is thus exploited
% whenever it happens to be available, at a small and bounded cost when it is not.
% An adapter matches the side information to the entropy model, and training over
% random side-information qualities gives one model for every quality. Delivering
% an image to a receiver that already holds the quality just below costs up to
% $46\%$ fewer bits, or $52\%$ with an extra hyper-latent, while the path without
% side information loses under $4\%$ and reconstructions stay bit-identical. A
% half-resolution first round recovers the most bits per bit sent.
% \end{abstract}

\begin{abstract} 
In many delivery settings, a receiver may already hold a lower-quality or lower-resolution representation of an image, obtained through an independent transmission. Conventional codecs encode a subsequently requested higher-quality representation without exploiting this incidental side information, whereas conditional codecs generally assume a prescribed source of side information that is always available. We instead consider an \emph{opportunistic} setting in which side information may or may not be present.
We introduce a single entropy model that conditions on a previously decoded latent when available and falls back to a standard hyperprior otherwise. The proposed adapter maps the side-information latent to the prior signal required by the entropy model, allowing the same model to support multiple target and side-information quality combinations. The analysis and synthesis transforms remain frozen, enabling retrofitting of an existing learned codec while preserving its latent representation and reconstruction path.
When the receiver holds the quality immediately below the target, the proposed method reduces the rate of the subsequent transmission by up to $46\%$, or by $52\%$ when an additional hyper-latent is transmitted. In the absence of side information, the rate penalty remains below $4\%$, and the reconstructions are bit-identical across the conditional and fallback modes. 
%Among the considered first-round representations, a half-resolution transmission provides the largest second-round rate saving per bit previously transmitted. 
\end{abstract}


\section{Introduction}
\label{sec:intro}

% Learned image compression now matches or exceeds classical codecs, and most
% competitive designs descend from the hyperprior family: a latent is coded under
% a Gaussian entropy model whose parameters are predicted from a transmitted
% hyper-latent~\cite{Balle2018ICLR,minnen2018joint}. As this technology is
% standardised in JPEG~AI~\cite{Alshina2024Multim}, the analysis and synthesis
% transforms become fixed at deployment. These are also the largest and most
% hardware-sensitive parts of a receiver, so an extension that requires
% redesigning or retraining them is difficult to deploy.

% Learned image compression now matches or outperforms conventional codecs under several distortion criteria. A typical neural codec comprises nonlinear analysis and synthesis transforms, together with an entropy model for lossless coding of the quantised latent representation. These components are commonly trained jointly end to end. The most competitive architectures descend from the hyperprior family, in which a latent produced by the analysis transform is coded under a Gaussian entropy model whose parameters are predicted from a transmitted hyper-latent~\cite{Balle2018ICLR,minnen2018joint}.

Learned image compression now matches or outperforms conventional codecs under several distortion criteria. A typical learned codec combines nonlinear analysis and synthesis transforms with an entropy model for losslessly coding the quantised latent representation. In the hyperprior architectures underlying many competitive and mainstream codecs, the analysis latent is modelled by a Gaussian distribution whose parameters are predicted from a transmitted hyper-latent~\cite{Balle2018ICLR,minnen2018joint}. These components are typically trained jointly end to end.

% As this technology is
% standardised in JPEG~AI~\cite{Alshina2024Multim}, the analysis and synthesis
% transforms become fixed at deployment. These are also the largest and most
% hardware-sensitive parts of a receiver, so an extension that requires
% redesigning or retraining them is difficult to deploy.
% As learned image compression progresses towards standardization, notably through JPEG~AI~\cite{Alshina2024Multim}, extensions that preserve existing codec components become desirable. Moreover, retraining either transform remains costly and may disrupt existing latent representations.
As learned codecs mature, including through standardisation efforts such as JPEG~AI~\cite{Alshina2024Multim}, extensions that preserve existing codec components become increasingly attractive. Retraining the analysis or synthesis transform can be costly and may alter the latent representation or reconstruction path. We therefore consider whether an existing codec can be extended through its entropy model alone.

% At the same time, many delivery settings leave a decoded version of the very
% same image at the receiver: a lower rung of an adaptive-streaming quality
% ladder, a cached preview or thumbnail, a resolution ladder, or a stream sent
% earlier for a machine-vision task. Such content is legitimate side information,
% and conditioning an entropy model on side information $S$ cannot increase the rate, since
% $H(Y\mid S)\le H(Y)$ with a gain bounded by the mutual information $I(Y;S)$.
% Conditional coding has been exploited in exactly this way for video and, more
% recently, for images~\cite{Li2021DCVC,Liu2020ECCV,Shen2024ISCAS}. Existing
% schemes, however, co-design the encoder, the decoder and the entropy model
% around one specific side-information source that is assumed always present.

% At the same time, a receiver may already hold a decoded representation of the same image, such as a lower rung of an adaptive-streaming quality ladder, a cached preview or thumbnail, a lower-resolution version, or a stream previously transmitted for a machine-vision task. Because this representation is available at both ends, it can serve as side information $S$ for entropy coding: conditioning the target latent $Y$ on $S$ can reduce its entropy from $H(Y)$ to $H(Y\mid S)$, with a potential gain of $I(Y;S)$~\cite{slepian1973tit}. Related conditional-coding principles have been exploited in video and learned image compression~\cite{Li2021DCVC,Liu2020ECCV,Shen2024ISCAS}. However, existing methods typically assume a prescribed side-information source and co-design the codec around its availability.

Our setting arises when the receiver already holds an independently decoded representation of the same image, such as a cached preview, a lower rung of a quality or resolution ladder, or a representation transmitted for a machine-vision task. Because the corresponding latent is available at both the encoder and decoder, it can serve as side information $S$ for coding the target latent $Y$. In principle, conditioning reduces the required rate from $H(Y)$ to $H(Y\mid S)$, with a potential gain of $I(Y;S)$~\cite{slepian1973tit}. Conditional coding has been studied in video and learned image compression~\cite{Li2021DCVC,Liu2020ECCV,Shen2024ISCAS}, but existing methods typically assume a prescribed source of side information and design the codec around its availability.

We instead freeze the analysis and synthesis transforms and retrain the entropy model once. We then ask: 
\begin{quote} \emph{How cheaply can a target image be transmitted when the receiver may already hold a lower-quality or lower-resolution representation of the same image?} 
\end{quote}
Unlike a designed base layer, this side information is \emph{opportunistic}: its availability and quality are not known at training time, and the codec must remain effective without it.

% We evaluate the rate of the \emph{subsequent} transmission rather than the aggregate rate. We do not claim a net gain over directly coding the target: in our experiments, the first-round bitstream costs more than it saves (Section~\ref{sec:firstround}). Our setting instead assumes that the earlier representation was required independently, for example as part of a quality or resolution ladder, a cache, or a machine-vision stream. Finally, the method does not preserve an unchanged decoder: it modifies and retrains the entropy model while leaving the transforms and reconstruction path unchanged. 
We focus on the rate of the \emph{subsequent} transmission, not the aggregate rate of both transmissions. Accordingly, we do not claim a net gain over directly coding the target image: in our experiments, the earlier bitstream costs more than it saves (Section~\ref{sec:firstround}). Rather, we assume that the earlier representation was required independently. We also do not preserve an unchanged decoder: the entropy model is modified and retrained, while the transforms and reconstruction path remain unchanged.
% \begin{itemize} 
% \item A \emph{dual-mode} entropy model that switches between an external, previously decoded latent and the standard hyperprior. Both modes share one fixed entropy model and the same frozen transforms, while producing bit-identical reconstructions. This enables conditional coding when side information is available, with only a small rate penalty when it is absent. \item Two constructions for the conditional prior: an adapter-only design that introduces no additional conditional hyper-latent, and a joint hyper-encoder design that provides a stronger prior by transmitting an additional hyper-latent. 
% \item A single model trained across side-information qualities that supports all tested quality combinations and generalizes to side-information resolutions not encountered during training, demonstrated on MBT-2018 and its variable-rate variant. 
% \end{itemize}
Our contributions are: 
\begin{itemize} 
\item A \emph{dual-mode} entropy model that switches between a previously decoded latent and the standard hyperprior. A single fixed model supports both modes, produces bit-identical reconstructions, and incurs only a small rate penalty when side information is unavailable. 
\item Two constructions for the conditional prior: an adapter-only design that requires no additional conditional hyper-latent, and a joint hyper-encoder design that transmits an additional hyper-latent to obtain a stronger prior.
\item A model trained jointly across side-information qualities that supports all tested quality combinations and generalises to side-information resolutions not seen during training, demonstrated with MBT-2018~\cite{minnen2018joint} and its variable-rate variant~\cite{Kamisli2024DCC}. 
\end{itemize}

\section{Related work}
\label{sec:related}

% \textbf{Hyperprior codecs.} The scale hyperprior~\cite{Balle2018ICLR} predicts
% per-coefficient scales from a transmitted hyper-latent; MBT-2018 adds means and
% an autoregressive context~\cite{minnen2018joint}. We build on the CompressAI
% implementations~\cite{begaint2020,CompressAIGithub} and on the variable-rate
% variant of~\cite{Kamisli2024DCC}. JPEG~AI standardises a learned
% codec~\cite{Alshina2024Multim}, which is what motivates leaving the transforms
% untouched.
\subsection{Hyperprior codecs} 
The scale hyperprior~\cite{Balle2018ICLR}, the first learned codec to introduce a hierarchical prior, predicts coefficient-wise scales from a transmitted hyper-latent. Its direct extension, MBT-2018~\cite{minnen2018joint}, additionally predicts means and incorporates an autoregressive context. More recently, Kamisli \emph{et al.} proposed a variable-rate extension of this architecture~\cite{Kamisli2024DCC}. JPEG~AI~\cite{Alshina2024Multim} standard also builds on the hyperprior paradigm, further motivating extensions that preserve established codec components. Accordingly, in our setting, the analysis and synthesis transforms remain frozen, and only the entropy model is retrained.

% \textbf{Conditional entropy coding.} DCVC showed that conditioning dominates
% residual coding for video~\cite{Li2021DCVC}, and conditional entropy models
% across frames are now standard~\cite{Liu2020ECCV}. For still images, Shen
% \emph{et al.} code a low-resolution image, super-resolve it, and condition the
% codec on the result~\cite{Shen2024ISCAS}; the main encoder, the hyperprior
% \emph{and} the synthesis transform are all rebuilt around that condition, which
% is always present. Scalable codecs such as DeepFGS~\cite{Zhai2025DCC} split the
% image into base and enhancement features and tie them with a mutual entropy
% model, but the base layer is part of one jointly trained bitstream. Our first
% round is instead an independent, standard-decodable bitstream, and only the
% entropy model is retrained.
\subsection{Conditional and scalable coding.} Conditional entropy models are widely used to exploit correlations between representations, following the information-theoretic principles of distributed source coding established by Slepian and Wolf~\cite{slepian1973tit}. In learned video compression, DCVC demonstrated the advantage of conditioning over explicit residual coding~\cite{Li2021DCVC}, building on conditional entropy models across frames~\cite{Liu2020ECCV}. Still for images, Shen \emph{et al.} code a low-resolution image, super-resolve it, and condition the target codec on the resulting reconstruction~\cite{Shen2024ISCAS}. Their encoder, hyperprior, and synthesis transform are jointly designed around side information that is always present. Scalable codecs such as DeepFGS~\cite{Zhai2025DCC} jointly train base and enhancement features connected by a mutual entropy model. JPEG~AI also supports progressive decoding: its hyperstream provides a preview, and ordered latent channels progressively refine the reconstruction~\cite{Alshina2024Multim}. These approaches construct a single codestream whose coarse representation is designed as part of the target coding process. By contrast, we exploit a previously and independently transmitted representation that was not designed as a base layer, may have different quality or resolution, and may be absent altogether.

\subsection{Absent side information.} That side information may or may not be
available is the classical Heegard--Berger setting~\cite{Heegard1985TIT}, the
information-theoretic counterpart of our switch. Closest in vocabulary,
CASH~\cite{Deniffel2026ICASSP} switches among several \emph{internal} hyperprior
networks per image; the switch never involves externally decoded content.
Conditional coding for human--machine streams~\cite{Tatsumi2025MMSP} shares our
motivation but redesigns the codec.

To our knowledge, no prior work keeps one entropy model, behind one fixed pair
of transforms, usable both with and without an external decoded latent of
\emph{arbitrary} quality.

\section{Method}
\label{sec:method}

\begin{figure*}[t]
\centering
\includegraphics[width=0.8\textwidth]{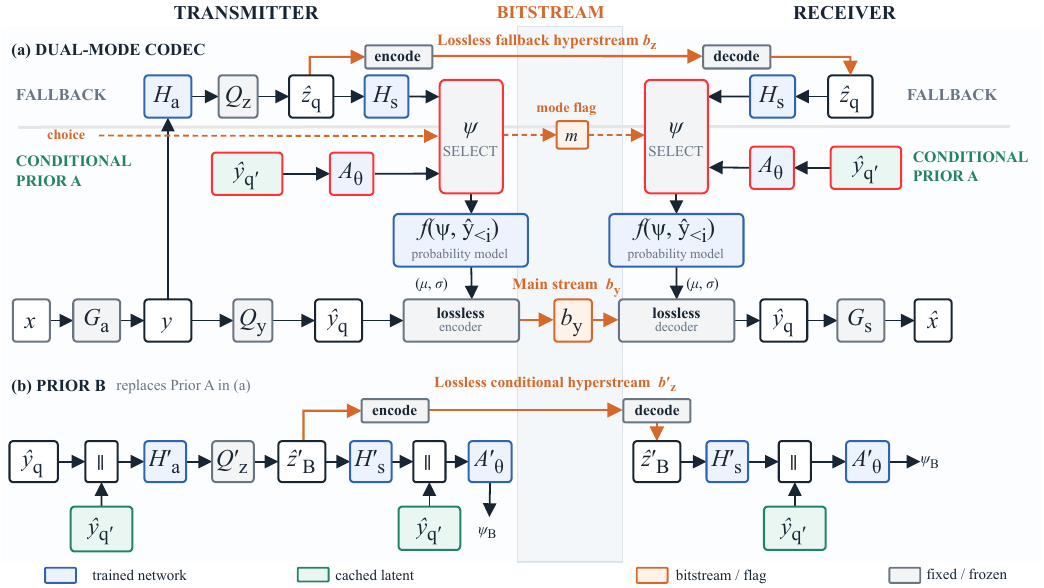}
\caption{Architecture of the proposed dual-mode codec. (a) The transmitted flag $m$ selects, at both transmitter and receiver, between the fallback hyperprior and Prior~A (highlighted in red), which adapts the cached latent $\yh_{q'}$. The selected prior $\psi$ parameterizes the entropy model used to code the target latent $\yh_q$. The streams $b_y$ and $b_z$ carry the target and fallback hyper-latents, respectively. (b) Prior~B replaces Prior~A and additionally transmits the conditional hyper-latent through $b'_z$.} 
%Blue denotes trained networks, green cached latents, orange transmitted information, and gray fixed or frozen components.}

\label{fig:arch}
\end{figure*}

\subsection{Hyperprior background}

An image $x\in\mathbb{R}^{H\times W\times C}$ is mapped by the analysis transform to a latent representation $y=G_a(x)$, quantized as $\yh=Q(y)$, and reconstructed by the synthesis transform as $\hat{x}=G_s(\yh)$. To entropy-code $\yh$, the entropy model $f$ predicts Gaussian parameters for each of its $N_y$ coefficients from a prior signal $\psi$ and the autoregressive context: $(\mu_i,\sigma_i)=f_i(\psi,\yh_{<i})$ for $i=1,\ldots,N_y$. In a standard hyperprior codec, the prior signal is derived from a quantized hyper-latent as $\zh=Q\!\left(H_a(y)\right)$ and $\psi=H_s(\zh)$, where $H_a$ and $H_s$ denote the hyper-encoder and hyper-decoder, respectively, and $Q(\cdot)$ denotes element-wise uniform quantization. Each latent coefficient is coded under the discretized Gaussian model 
\begin{equation} 
\label{eq:gauss} 
p(\yh_i\mid\psi,\yh_{<i}) = \int_{\yh_i-\frac{1}{2}}^{\yh_i+\frac{1}{2}} \mathcal{N}\!\left(t;\mu_i,\sigma_i^2\right)\,\mathrm{d}t, 
\end{equation} yielding the total rate $R_{\mathrm{base}} = R(\yh\mid\psi)+R(\zh)$
with
\begin{equation} 
\label{eq:rate}
R(\yh\mid\psi) = -\sum_{i=1}^{N_y} \log_2 p(\yh_i\mid\psi,\yh_{<i})+R(\zh). 
\end{equation}

\subsection{A dual-mode entropy model}

In the standard hyperprior framework, $\psi$ is the interface through which the hyperprior informs the entropy model about the distribution of $\yh$. We therefore leave $f$, $G_a$, and $G_s$ structurally unchanged and modify only how $\psi$ is formed to exploit opportunistic side information (Fig.~\ref{fig:arch}): \begin{equation} \label{eq:switch} \psi= \begin{cases} 
\psi_{\mathrm{fb}} = H_s(\zh_q), & \text{fallback mode,}\\
\psi_{\mathrm{cond}} = A_\theta(\yh_{q'}), & \text{conditional mode,} 
\end{cases} 
\end{equation} where $\yh^{q'}$ is the latent of a previously transmitted representation of the same image at quality $q'\leq q$, already available at the receiver. The fallback mode, where no side information is used, is equivalent to the base mode. The convolutional adapter $A_\theta$ maps $\yh_{q'}$ to the prior-signal interface expected by $f$. The selected mode is signaled once per image. Therefore, the decoder does not need to infer whether side information is available. Specialising $R_{\mathrm{base}}$ according to the two branches of \eqref{eq:switch}, the rate expressions at target quality $q$ are
\begin{equation} 
\label{eq:rates} 
R_{\mathrm{fb}} = R\!\left(\yh_q\mid \psi_{\mathrm{fb}} \right)+R(\zh_q), \qquad R_{\mathrm{cond}} = R\!\left(\yh_q\mid \psi_{\mathrm{cond}} \right). \end{equation} 
In conditional mode, $A_\theta(\yh_{q'})$ replaces the hyper-decoder output $H_s(\zh_q)$ as the prior signal. The target hyper-latent $\zh_q$ is therefore not transmitted, and $R_{\mathrm{cond}}$ gives the complete rate of the subsequent bitstream, up to the mode-signalling overhead.

% \textbf{Reconstructions are bit-identical.} We quantise the latent directly,
% $\yh=Q(y)$, rather than the residual $Q(y-\mu)+\mu$ used by default in
% CompressAI. This is possible because the transforms are frozen and no
% rate--distortion trade-off has to be balanced, and it makes $\yh$ independent of
% $\mu$ and hence of $\psi$. Since $G_s$ is shared and frozen, the reconstruction
% $\hat{x}$ is then \emph{identical}, sample for sample, whether or not side
% information was used. Conditional coding here is purely a rate operation: the
% entropy model may be retrained, but no receiver ever produces a different
% picture, and distortion never enters the comparison.

\subsubsection{Bit-identical reconstructions.} 
In common practical implementations~\cite{begaint2020, CompressAIGithub}, quantisation is applied to the latent residual relative to the predicted mean, yielding $\hat y =  Q(y-\mu)  + \mu$. Consequently, changing the prior signal $\psi$ may change $\mu$ and, in turn, the reconstructed latent $\hat y$. We instead quantize the latent directly as $\yh=Q(y)$, making the reconstructed latent independent of $\psi$ and $\mu$. The entropy model therefore determines only the probabilities assigned to the quantized symbols, and hence their code lengths. Since both modes decode the same $\yh$ and apply the same frozen synthesis transform $G_s$, they produce exactly the same reconstruction $\hat{x}$. Thus, the conditional and fallback modes differ only in rate, not in the decoded image or its distortion.

% \subsection{Prior construction}
% \label{sec:priors}

% \textbf{Prior A (adapter only).} $\psi=A_\theta(\yh_{s'})$, with $A_\theta$ a
% $5\times5$ convolution (stride 1, padding 2), a LeakyReLU, and a $3\times3$
% convolution (stride 1, padding 1), both emitting the $2M$ channels the entropy
% model expects, with $M$ the number of latent channels. Nothing is added to the bitstream. Because $A_\theta$ takes the
% place of $H_s$ rather than sitting alongside it, the conditional path does not
% grow the decoder \pending{report parameters of $A_\theta$ against $H_s$}. No
% explicit signal of $s'$ is provided to the model.

% \textbf{Prior B (joint hyper-encoder).} A stronger prior is obtained by letting
% a new hyper-encoder see the target latent as well:
% \begin{equation}
% \label{eq:priorb}
% \zh'=Q\!\left(H'_a(\yh\cat\yh_{s'})\right),
% \qquad
% \psi=A'_\theta\!\left(H'_s(\zh')\cat\yh_{s'}\right),
% \end{equation}
% where $\cat$ is channel concatenation and $H'_a$ mirrors the MBT hyper-encoder.
% The concatenation uses the quantised latents, matching what the decoder will
% hold.
% Reusing $\yh_{s'}$ at both ends of the hyper-encoder follows~\cite{Wang2020AAAI}.
% Prior B must transmit $\zh'$ and requires $H'_s$ at the receiver, so unlike
% Prior A it does change the transmitted syntax; we report it as an upper bound on
% what the conditioning signal can deliver.

\subsection{Prior construction} 
\label{sec:priors} 
We consider two constructions for deriving the prior signal $\psi$ from the available side-information latent $\yh_{q'}$. 

\subsubsection{Prior A (adapter only).} The prior signal is obtained directly from the side-information latent as $\psi=A_\theta(\yh_{q'})$, where $A_\theta$ consists of a $5\times5$ convolution with stride~1 and padding~2, followed by a LeakyReLU and a $3\times3$ convolution with stride~1 and padding~1. Its output has $2M$ channels, matching the input expected by the entropy model, where $M$ denotes the number of main-latent channels. In conditional mode, $A_\theta$ replaces the conventional hyper-decoder $H_s$ rather than operating alongside it. Consequently, no additional hyper-latent is transmitted and the conditional rate is simply $R(\yh\mid\psi)$. This construction therefore leaves the bitstream payload unchanged, although the decoder must support the adapter and know when the conditional mode is used. The side-information quality $q'$ is not explicitly provided to the model.

\subsubsection{Prior B (joint hyper-encoder).} 
A more expressive prior is obtained by allowing a new hyper-encoder to observe both the target and side-information latents: 
\begin{equation} 
\label{eq:priorb} \zh'_{\mathrm{B}} = Q\!\left(H'_a(\yh\cat\yh_{q'})\right), \qquad \psi_{\mathrm{B}} = A'_\theta\!\left(H'_s(\zh')\cat\yh_{q'}\right), 
\end{equation} 
where $\cat$ denotes channel-wise concatenation, and $H'_a$ and $H'_s$ follow the architectures of the MBT hyper-encoder and hyper-decoder, respectively. Both inputs are quantized latents, ensuring that the prior is constructed from representations available identically at the encoder and decoder. Conditioning the hyperprior on $\yh_{q'}$ at both the hyper-encoder and hyper-decoder follows the construction of~\cite{Wang2020AAAI}. Unlike Prior~A, Prior~B requires transmitting $\zh'$ and deploying the corresponding hyper-decoder $H'_s$ at the receiver. It therefore changes the conditional bitstream syntax and is included as a higher-complexity reference for assessing how effectively the available side information can reduce the coding rate.

% \subsection{Priors at a different resolution}

% If the first round was sent at a lower resolution, $\yh_{s'}$ does not match the
% latent grid. For half-resolution side information we prepend a $5\times5$
% transposed convolution with $M$ output channels; for three-quarter resolution, a
% stride-4 transposed convolution followed by a stride-3 convolution reaches the
% target grid, after which $5\times5$ and $3\times3$ layers produce $2M$ channels.
% Nothing else changes, and the entropy model is unaware of the side-information
% geometry.

\subsection{Priors of a different resolution}
When the previously transmitted representation has a lower resolution than the target, the side-information latent $\yh_{q'}$ and target latent $\yh_q$ have different spatial dimensions. The adapter therefore first maps $\yh_{q'}$ to the target latent grid. For half-resolution side information, we use a $5\times5$ transposed convolution with stride~2 and $M$ output channels. For three-quarter-resolution side information, we use a stride-4 transposed convolution followed by a stride-3 convolution, yielding the required $4/3$ spatial scaling. The resulting features are then processed by the standard $5\times5$ and $3\times3$ adapter layers to produce the $2M$-channel prior signal $\psi$. Thus, the resolution-dependent processing is confined to the adapter, while the entropy model remains unchanged.

% \subsection{Training}

% $G_a$ and $G_s$ are frozen; only the adapters $A_\theta$ and $A'_\theta$ and the
% entropy model are updated. Each batch draws a side-information quality $s'\le s$ and, on a
% fraction of the batches, the fallback branch of~\eqref{eq:switch} instead
% \pending{confirm the sampling schedule of the fallback branch}, so one model
% learns to operate with any amount of side information and with none. For the
% variable-rate codec the target quality $s$ is drawn per batch as well, and a
% single trained entropy model therefore covers every $(s,s')$ pair we report
% %\pending{confirm that the MBT-VBR results at $s\!=\!7$ and $s\!=\!5$ come from one checkpoint}. 
% Since the reconstruction path is fixed, distortion is constant and
% the objective reduces to the rate,
% \begin{equation}
% \label{eq:loss}
% \mathcal{L}=R(\yh\mid\psi)\;(+\,R(\zh')\ \text{for Prior B}).
% \end{equation}
% Consequently PSNR is the same for every configuration at a given $s$
% (36.93~dB at MBT-2018 $s\!=\!8$) and rates can be compared directly, without
% rate--distortion curves.

\subsection{Training} We freeze the analysis and synthesis transforms $G_a$ and $G_s$ and update only the entropy model $f$ and the adapters $A_\theta$ and $A'_\theta$. For each training batch, we sample a target quality $q$ and either a side-information quality $q'\leq q$ or the fallback mode with no side information. 
%\pending{Specify the probability of sampling the fallback mode.} 
This exposes the same entropy model to multiple side-information qualities and to their absence. For the variable-rate codec, sampling $q$ as well as $q'$ allows a single checkpoint to cover all reported $(q,q')$ combinations. Because the reconstruction path is frozen, training optimizes rate alone. For Prior~A, the loss for each batch is selected according to the active mode as defined in \eqref{eq:rates}.
For Prior~B, the conditional loss additionally includes the rate of the transmitted hyper-latent: 
\begin{equation} 
\label{eq:loss-prior-b}
\mathcal{L}_{\mathrm{B}} = R\!\left(\yh_q\mid\psi_{\mathrm{B}}\right) + R\!\left(\zh'_{\mathrm{B}}\right).
\end{equation}
Since neither the target latent $\yh_q$ nor the synthesis transform changes during training, every configuration at a given target quality produces the same reconstruction and hence the same distortion. Rates can therefore be compared directly at fixed PSNR, without plotting rate--distortion curves.
% For example, all MBT-2018 configurations at $q=8$ yield an average PSNR of $36.93$~dB.

\section{Numerical experiments}
\label{sec:experiments}

\subsection{Simulation settings}

% We use the CompressAI~\cite{begaint2020,CompressAIGithub} MBT-2018 model and the
% variable-rate MBT-2018-VBR of~\cite{Kamisli2024DCC}, initialised from the
% pretrained weights. The two differ in a way that matters here: MBT-2018 reaches
% its eight qualities with eight separately trained checkpoints, so $\yh_{s'}$ and
% $\yh$ come from different analysis transforms, whereas MBT-2018-VBR covers all
% eight with one transform and a quality offset, which keeps every $\yh_{s'}$ in
% the latent space of the target. Training and evaluation use COCO~\cite{lin2014microsoft},
% with $118{,}287$ images for training ($256\times256$ random crops, horizontal
% flip and colour jitter) and $40{,}670$ for evaluation ($512\times512$ crops).
% The entropy model is trained for 40 epochs with Adam, learning rate $10^{-4}$
% and batch size 16. Downscaled side information uses a Lanczos filter. The indices $s$ and
% $s'$ enumerate the eight CompressAI quality points, numbered $1\dots8$ for
% MBT-2018 and $0\dots7$ for the variable-rate model.

% Note that $s'\!=\!s$ is degenerate: the conditioning latent is then the target
% latent itself, so the receiver already holds the image and nothing needs to be
% delivered. We report that column only as an upper bound on what conditioning can
% achieve, and quote all headline figures at $s'\!=\!s-1$.

We evaluate the proposed method using the CompressAI MBT-2018 model~\cite{begaint2020,CompressAIGithub} and the variable-rate MBT-2018-VBR model proposed in 2024~\cite{Kamisli2024DCC}, both initialized from pretrained weights. MBT-2018 represents its eight quality levels with independently trained checkpoints; hence, $\yh_{q'}$ and $\yh_q$ may be produced by different analysis transforms and need not be aligned. In contrast, MBT-2018-VBR uses a single analysis transform controlled by a quality-dependent offset, producing both latents within a shared representation that may facilitate prediction of $\yh_q$ from $\yh_{q'}$.

We use $118{,}287$ COCO images for training and $40{,}670$ for evaluation~\cite{lin2014microsoft}. Training uses random $256\times256$ crops, horizontal flipping, and color jitter, whereas evaluation uses $512\times512$ crops. We train the entropy model for 40 epochs with Adam, a learning rate of $10^{-4}$, and a batch size of 16. Downscaled side information is generated using Lanczos filtering. The quality indices $q$ and $q'$ range from $1$ to $8$ for MBT-2018 and from $0$ to $7$ for MBT-2018-VBR.

When $q'=q$, the conditioning and target latents are identical, so the receiver already possesses the target representation and the ideal incremental rate is zero. We retain this case as a diagnostic of the entropy model under maximally informative side information, while reporting headline results for the practical setting $q'=q-1$.

% Table I removed: the MBT-VBR s=7 and MBT-2018 s=8 rows are all readable off
% Fig. 2 (curves at s'=s-1 and s'=s, plus the two fallback lines). Only the
% s=5 row was unique to it, and those numbers now live in the prose of
% Sec. IV-B and IV-C. Kept here in case a reviewer asks for the tabulated form.
%
% \begin{table}[t]
% \centering
% \caption{Rate of the bitstream that delivers the image, in bpp; the hyper-latent
% ($0.013$~bpp) is excluded throughout and is sent only in fallback mode. The
% $s'\!=\!s$ columns are the degenerate case discussed in the text and bound what
% conditioning can achieve. Reconstructions, and hence PSNR, are identical across
% each row.}
% \label{tab:main}
% \scriptsize
% \setlength{\tabcolsep}{4pt}
% \begin{tabular}{lcccccc}
% \toprule
%  & \multicolumn{2}{c}{fallback (no SI)} & \multicolumn{2}{c}{$s'\!=\!s-1$}
%  & \multicolumn{2}{c}{$s'\!=\!s$ (bound)} \\
% \cmidrule(lr){2-3}\cmidrule(lr){4-5}\cmidrule(lr){6-7}
% Codec & pretr. & ours & Prior~A & Prior~B & Prior~A & Prior~B \\
% \midrule
% MBT-VBR, $s\!=\!7$  & 1.782 & 1.843 & 1.034 & 0.885 & 0.929 & 0.707 \\
% MBT-VBR, $s\!=\!5$  & 1.143 & 1.179 & 0.629 & 0.566 & 0.470 & 0.350 \\
% MBT-2018, $s\!=\!8$ & 1.765 & 1.798 & 1.270 & 1.045 & 0.928 & 0.400 \\
% \bottomrule
% \end{tabular}
% \end{table}

\begin{figure*}[t]
\centering
\includegraphics[width=0.8\textwidth]{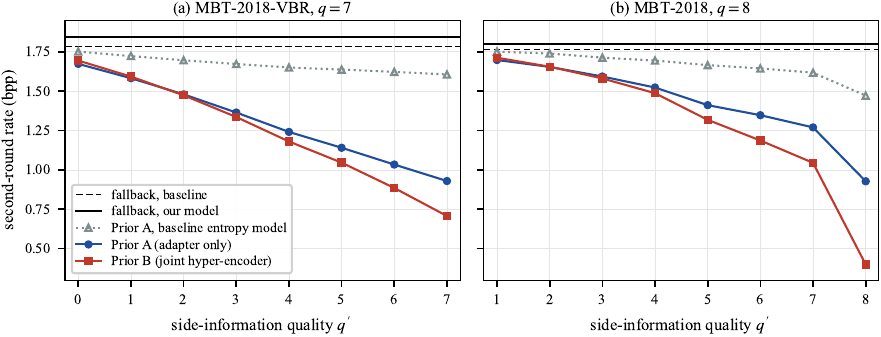}
\caption{Second-round rate as a function of the side-information quality $q'$.
The curves are the complete conditional rate $R_{\mathrm{cond}}$ (no target
hyper-latent); the horizontal lines are fallback and baseline second-round
rates, which additionally carry $\zh$ ($0.013$~bpp at MBT-2018 $q\!=\!8$).
The gap between the horizontal lines is the price of dual-mode operation.
The rightmost point of each curve is the degenerate $q'\!=\!q$ case.
The first-round rate is not included; see Section~\ref{sec:firstround}.}
\label{fig:main}
\end{figure*}

\subsection{Second-round performance}

The second-round rate is the cost of delivering the target $\yh_q$: $R_{\mathrm{cond}}$ when a previously decoded latent is used in conditional mode, and $R_{\mathrm{fb}}$ when the model falls back to the hyperprior. Figure~\ref{fig:main} reports both. The first-round rate that produced the side information is excluded and is discussed in Section~\ref{sec:firstround}. 

As expected, under both Prior~A and Prior~B, $R_{\mathrm{cond}}$ decreases as $q'$ increases because the receiver has more informative side information.
For MBT-2018-VBR (Fig.~\ref{fig:main}(a)), when the target quality is $q=7$ and the receiver holds the same image at the immediately preceding quality, $q'=6$, Prior~A reduces the second-round rate to $1.034$~bpp, $43.9\%$ below fallback. At $(q,q')=(5,4)$ (not shown in the figure), it requires $0.629$~bpp, corresponding to a $46.6\%$ reduction. Prior~A achieves these savings without transmitting a target hyper-latent. Prior~B transmits the additional hyper-latent $\zh'$ and increases the reduction to $52.0\%$ in both cases.

The fixed-rate MBT-2018 model (Fig.~\ref{fig:main}(b)) benefits less uniformly from side information. At $q'=7$, Prior~A and Prior~B reduce its second-round rate by $29.4\%$ and $41.9\%$, respectively. Although its rate ladder is similar to that of MBT-VBR, the major gain occurs in the final step, $q'=q$, when the side-information and target latents are produced by the same quality-specific checkpoint. This is consistent with independently trained MBT-2018 checkpoints producing less aligned representations across qualities than the shared variable-rate model.

% Pushing to the degenerate $s'\!=\!s$ bounds what conditioning could ever
% achieve: $-48.4\%$ for Prior~A and $-77.8\%$ for Prior~B on MBT-2018, and
% $-60.1\%$ and $-70.3\%$ on MBT-VBR at $s\!=\!5$. That the
% bound is not $-100\%$ is itself informative, since $\yh_{s'}$ is then the target
% latent exactly; the residual cost measures how well the entropy model converts
% perfect side information into a distribution.\footnote{Setting the mean equal to
% $\yh_{s'}$ and forcing $\sigma$ close to zero does drive the rate to nearly
% zero, so the parameterisation admits an almost free solution at $s'\!=\!s$; the
% rate objective does not converge to it. That solution is confined to the
% degenerate case: once $s'\!<\!s$ the mean is inexact and a near-zero $\sigma$
% inflates the rate instead. We therefore never constrain $\sigma$ and let the
% entropy model learn it as it does without side information.} 

% The gap between the two horizontal lines in each panel of Fig.~\ref{fig:main}
% isolates what dual-mode capability costs. On the fallback path our entropy model
% spends $1.9\%$ more than the baseline on MBT-2018 ($1.765$ against
% $1.798$~bpp) and $3.4\%$ more on MBT-VBR $s\!=\!7$ ($1.782$ against
% $1.843$~bpp), with $3.1\%$ at $s\!=\!5$ ($1.143$ against $1.179$~bpp);
% including the hyper-latent, MBT-2018 goes from $1.778$ to $1.811$~bpp. A
% receiver therefore gains opportunistic conditional decoding for under $4\%$ on
% the fallback path, and never loses the ability to decode a stream sent without
% side information.
The two horizontal lines in each panel of Fig.~\ref{fig:main} are second-round rates in the absence of side information: the original codec $R_{\mathrm{base}}$ versus our fallback model $R_{\mathrm{fb}}$. Their difference measures the cost of supporting conditional operation when no side information is available. For MBT-2018, the main-latent rate increases from $1.765$~bpp for the original codec to $1.798$~bpp for our fallback model, a penalty of $1.9\%$. Including the hyper-latent, the total rate increases from $1.778$ to $1.811$~bpp. For MBT-VBR, the fallback rate increases from $1.782$ to $1.843$~bpp at $q=7$ and from $1.143$ to $1.179$~bpp at $q=5$ (not shown in the figures), corresponding to penalties of $3.4\%$ and $3.1\%$, respectively. The dual-mode model therefore retains ordinary coding without side information at a rate penalty below $4\%$ in all reported configurations.

\subsection{Ablation studies}

\paragraph{Effect of retraining the entropy model.}
The first dotted curve in Fig.~\ref{fig:main} isolates the effect of retraining $f$. It uses the adapted side information through the same prior interface while keeping the baseline entropy model fixed. For MBT-VBR at $(q,q')=(7,6)$, this variant requires $1.606$~bpp, only $9.9\%$ below its fallback rate. Retraining the entropy model reduces the rate to $1.034$~bpp, or $43.9\%$ below fallback. Thus, adapting the external latent alone is insufficient; the entropy model must learn how to interpret the resulting prior.   

\paragraph{Are quality indices needed?}
A second ablation (not shown in the figures) tests whether the entropy model should explicitly receive $(q,q')$. Starting from Prior~A, a small network uses these indices to rescale the predicted $\sigma$. For MBT-VBR at $q=7$, this variant badly affects the rate by requiring $1.100$ rather than $1.034$~bpp at $q'=6$, and $1.007$ rather than $0.929$~bpp at $q'=7$. We therefore find no benefit from explicitly conditioning on $(q,q')$ and use a single model for all $q'$ (although the higher rate is probably attributable to poor optimization).

%The difference is below $0.01$~bpp for the weakest side information and increases as $q'$ approaches $q$; the same ordering holds at $q=5$ and for MBT-2018. 
%Since the index-aware model contains Prior~A as a special case, 

\paragraph{Perfect-side-information diagnostic.}
We evaluate the degenerate case $q'=q$, for which the side-information and target latents are identical, $\yh_{q'}=\yh_q$. Although not a practical transmission setting, since the receiver already holds the target representation, it measures how effectively the entropy model exploits perfect side information. Prior~A and Prior~B reduce the MBT-2018 rate by $48.4\%$ and $77.8\%$, respectively. For MBT-VBR at $q=5$, the corresponding reductions are $60.1\%$ and $70.3\%$. An ideal conditional model would assign nearly unit probability to each known target symbol and require almost no rate. The residual rate therefore reflects an optimisation or modelling gap rather than uncertainty about the target latent.\footnote{Setting the predicted mean to $\yh_{q'}$ and taking $\sigma$ close to zero drives the rate near zero at $q'=q$, confirming that the parameterisation admits such a solution, although training does not converge to it. For $q'<q$, this construction assigns very low probability to symbols that differ from the side information. We therefore learn $\sigma$ without this constraint.}

\begin{figure}[t]
\centering
\includegraphics[width=0.8\columnwidth]{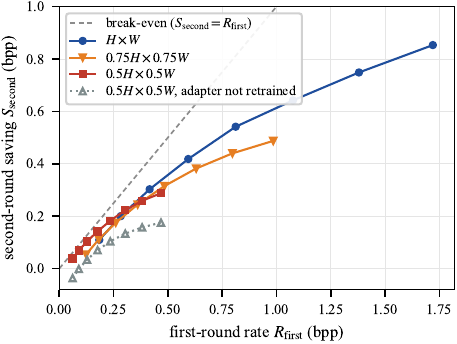}
\caption{Is it better to send a low-quality full-resolution image first, or a
higher-quality downscaled one? Saving on the second round against the rate of
the first round, MBT-VBR $q\!=\!7$, each curve traced over $q'$. Points below
the break-even line spend more on the first round than they recover.}
\label{fig:firstround}
\end{figure}

\subsection{First-round utility}
\label{sec:firstround}

Figure~\ref{fig:firstround} examines how to allocate a fixed first-round bit budget. In all experiments, the second round transmits the full-resolution target at $q=7$, whereas the first round uses full, three-quarter, or half resolution, with each curve traced by varying $q'$. The horizontal axis shows the rate $R_{\mathrm{first}}$ of the independently decodable first-round bitstream, and the vertical axis shows the resulting second-round saving, $S_{\mathrm{second}}=R_{\mathrm{fb}}-R_{\mathrm{cond}}$. The diagonal represents break-even operation, where $S_{\mathrm{second}}=R_{\mathrm{first}}$.

All evaluated points lie below this line, indicating that the total two-round cost exceeds that of direct fallback coding. Thus, if only the target image is required, omitting the first round is optimal. Accordingly, we now assume that a first-round representation is independently required and determine which resolution provides the greatest subsequent saving for a given budget.

Under this assumption, half-resolution side information performs best below approximately $0.33$~bpp. The half- and full-resolution curves intersect near $(R_{\mathrm{first}},S_{\mathrm{second}})=(0.33,0.23)$~bpp, above which full resolution performs better. The choice is not critical near the intersection: within $0.03$~bpp of the crossing, the difference remains below $0.01$~bpp. The half-resolution curve ends near $0.47$~bpp because no higher-rate operating point is available among the evaluated qualities. Across their respective ranges, $S_{\mathrm{second}}/R_{\mathrm{first}}$ is $61$--$81\%$ for half resolution, $50$--$73\%$ for full resolution, and $44$--$68\%$ for three-quarter resolution. 
%For example, half-resolution side information at $q'=7$ reduces the second-round rate to $1.495$~bpp, $17.3\%$ below fallback.

The half-resolution results also test generalization to an unseen side-information geometry, since the original adapter was not trained at this resolution. Even without retraining, the conditional model requires $1.607$~bpp, compared with $1.707$~bpp for the baseline entropy model (not shown in the figure), a $5.9\%$ reduction. Retraining the adapter for half-resolution inputs lowers the rate to $1.495$~bpp ($7.0\%$ below the unretrained configuration). These results show that the model generalizes to an unseen resolution, while geometry-specific training provides an additional gain. 
%The $1.707$~bpp baseline is omitted from Fig.~\ref{fig:firstround} because the vertical axis measures savings relative to fallback operation without side information.

\section{Conclusion}
\label{sec:conclusion}

% A single entropy model, behind one frozen pair of transforms, can serve both
% ordinary and opportunistic conditional decoding while producing bit-identical
% reconstructions. Delivering an image to a receiver that holds the quality one
% step below costs up to $46\%$ fewer bits without touching the transmitted
% syntax, or up to $52\%$ if an extra hyper-latent is allowed, while the
% fallback path loses under $4\%$; a half-resolution first round is the most
% bit-efficient. Two limitations bound the result: only about half of the
% first-round bits are recovered, so the scheme targets delivery settings in which
% that round is independently required, and even perfect side information leaves a
% substantial rate. The latter suggests that the entropy model's predictors of
% mean and variance, rather than the conditioning signal, are what limit the
% approach.

We presented a dual-mode codec in which a single entropy model and a frozen analysis--synthesis transform pair support both ordinary fallback and opportunistic conditional coding while preserving bit-identical reconstructions. When the receiver holds the same image one quality level below the target, Prior~A reduces the second-round rate by up to $46\%$ without transmitting an additional hyper-latent, while Prior~B achieves savings of up to $52\%$ by transmitting one. The fallback overhead remains below $4\%$, showing that conditional support can be added at little cost to ordinary operation. Additional training optimization is likely to improve these numbers.

Nevertheless, the first-round cost is never fully recovered through second-round savings. The method is therefore most relevant when an earlier representation is independently required. In this setting, half-resolution side information provides the best return at low first-round rates. Finally, the substantial residual rate under perfect side information further indicates that entropy-model optimisation remains the main bottleneck.
%, particularly the prediction of distribution parameters, 

\bibliographystyle{IEEEtran}
\bibliography{refs_abrv,refs_icassp}

\end{document}